\documentclass[12pt]{iopart}

\usepackage{iopams}  
\usepackage{graphicx}

\usepackage{caption}
\usepackage{subfig}

\usepackage{siunitx}

\usepackage[version=3]{mhchem}

\begin{document}

\title{Single-molecule study for a graphene-based nano-position sensor}

\author{G. Mazzamuto$^{*1,2}$, A. Tabani$^{*2}$, S. Pazzagli$^{2}$, S. Rizvi$^{1}$, A. Reserbat-Plantey$^{5}$, K. Sch\"{a}dler$^{5}$, G. Navickait\'e$^{5}$, L. Gaudreau$^{5}$, F.S. Cataliotti$^{1,2,3}$, F. Koppens$^{5}$, and C. Toninelli$^{1,3,4}$}

\address{$*$These authors equally contributed to this work}
\address{$^{1}$LENS and Universit\`{a} di Firenze, Via Nello Carrara 1, 50019 Sesto Fiorentino, Italy}
\address{$^{2}$Dipartimento di Fisica ed Astronomia, Via Sansone 1, 50019 Sesto Fiorentino, Italy}
\address{$^{3}$QSTAR, Largo Enrico Fermi 2, 50125 Firenze, Italy}

\address{$^{4}$INO, Istituto Nazionale di Ottica, Largo Fermi 6, 50125 Firenze, Italy}
\address{$^{5}$ICFO-Institut de Ciencies Fotoniques, Mediterranean Technology Park, 08860 Castelldefels, Barcelona, Spain}

\ead{toninelli@lens.unifi.it}
\begin{abstract}
In this study we lay the groundwork for a graphene-based fundamental ruler at the nanoscale. It relies on the efficient energy-transfer mechanism between single quantum emitters and low-doped graphene monolayers. Our experiments, conducted with dibenzoterrylene (DBT) molecules, allow going beyond ensemble analysis due to the emitter photo-stability and brightness. A quantitative characterization of the fluorescence decay-rate modification is presented and compared to a simple model, showing agreement with the $d^{-4}$ dependence, a genuine manifestation of a dipole interacting with a 2D material. With DBT molecules, we can estimate a potential uncertainty in position measurements as low as \SI{5}{\nm} in the range below \SI{30}{\nm}.

\end{abstract}

\tableofcontents

\maketitle
\renewcommand{\leftmark}{}

\section{Introduction: single emitters close to interfaces}

A single atom or molecule in the vicinity of a planar layered medium represents a paradigmatic system in near field physics \cite{Novotny1997}. On-chip antenna design e.g., as well as surface-enhanced spectroscopy \cite{Moskovits1985}, require a quantitative analysis of the dipole radiation and energy distribution for such geometries.

This study is then essential in the quest for efficient light-matter interfaces, on which  optical sensing and photon-based communication protocols strongly rely \cite{Homola1999,OBrien2009}.  

Experiments performed by Drexhage in 1970 \cite{Drexhage1970} already showed a clear modification of emitter decay rate and radiation pattern as a function of the distance to a reflecting interface. Classically, this is understood as a feedback effect of the reflected complex field on the molecule itself. Quantum mechanically, vacuum fluctuations depend on the problem boundary conditions and affect the density of states. The radiative decay rate results modified, as well as the efficiency of all those processes which can be described by the exchange of virtual photons. This is the case e.g. of a fundamental process in nature, i.e.\ photosynthesis, relying on the energy transfer between different chromophores. 

Molecules close to surfaces are also ideal probes of local effects and fields, as they are affected by the local environment at the nm scale, i.e.\ on the order of their physical size \cite{Moerner1999, Michaelis1999, Veerman1999}. Furthermore, similarly to what happens in cavity quantum electrodynamics, strong interactions of quantum emitters (QE) or absorbers with guided modes at the interface between different media can lead to cooperative effects \cite{Martin-Cano2010} and entanglement phenomena \cite{tudela2011, tudela2014}. 

A QE placed in the vicinity of a surface with complex electric permittivity can decay into three different channels \cite{Amos1997}: excitation of surface plasmons that propagate along the surface, radiation of photons into the far field, and dissipation through Ohmic losses. The absolute and relative probability of these processes depends on the specifications of the investigated system. Among other materials, graphene is now very well known for its unique optical, electronic and mechanical properties, which are a result of its gapless band structure and locally linear dispersion relation \cite{Neto2009}. From a fundamental point of view, the effects occurring for the case of a quantum emitter in close proximity to such a purely two-dimensional material are still largely unexplored, although they have the potential to set a new scenario for the physics of strong light-matter interaction \cite{Koppens2011}. In a broad frequency range, i.e.\ for energies above twice the Fermi energy and  for distances within a quarter of the wavelength, the main mechanism of energy relaxation is non-radiative dipole-dipole resonant energy transfer \cite{Chen2010}. In particular, graphene's unequaled conductivity gives rise to a very strong polarizability, resulting in an efficient energy transfer from donor molecules. As a consequence, graphene's use in the realm of functional materials, e.g.\ as an extraordinary energy sink for photodetection, seems advantageous. Based on the near-field interaction between biomolecules and graphene, a wealth of sensing applications have been proposed \cite{Zhang2011, Wang2011g}. More recently, even quantum position sensors have been suggested, relying e.g.\ on the Casimir effect exerted by graphene on a nearby two level system \cite{Muschik2013}. Similar quantum technologies have been discussed in the literature with the purpose of detecting and manipulating mechanical degrees of freedom of nano-oscillators \cite{Puller2013, Arcizet2011}.

In this letter we provide a first proof of principle for a graphene nano-ruler, based on the measurement of the energy-transfer rate between single organic molecules and a graphene monolayer. 

\section{Nanoscopic rulers by optical means}
\begin{figure}
\begin{center}
\includegraphics[width=\textwidth]{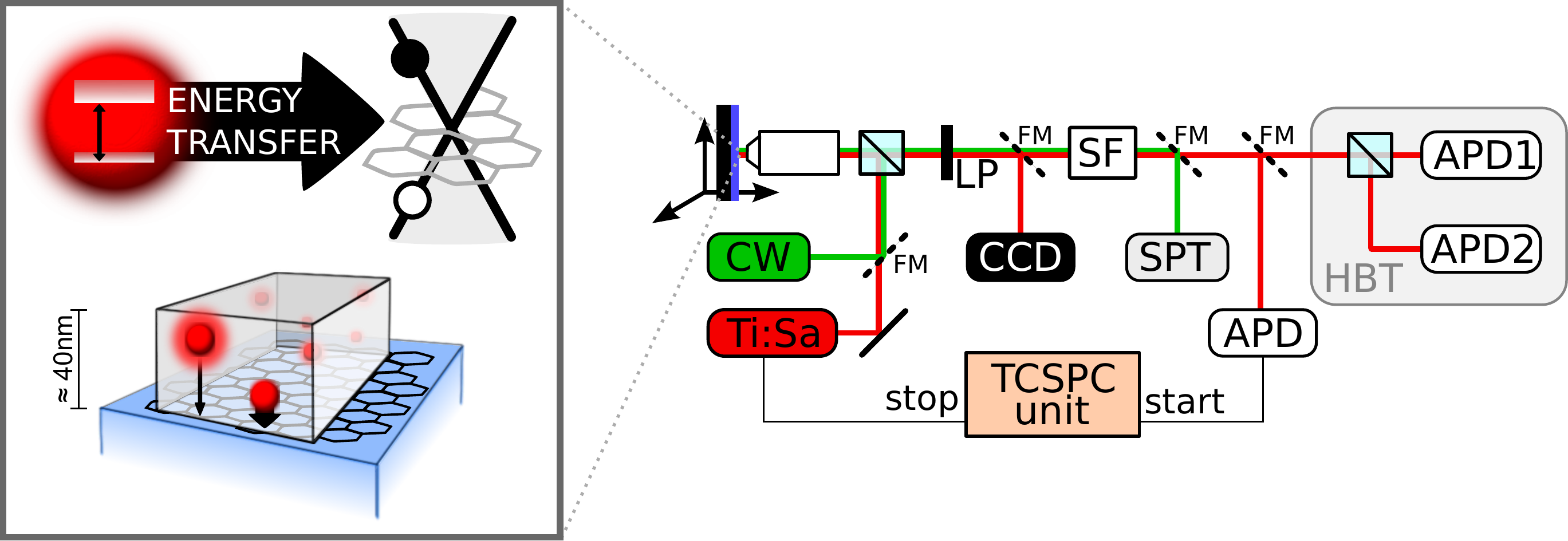}
\caption{(a) Artistic picture of our test sample, where a DBT:anthracene crystal is interfaced with a graphene monolayer sheet. Depending on the relative distance between molecule and graphene, the excited state decays with different branching ratios via radiative emission or energy transfer to graphene excitons. (b) Sketch of the experimental setup, combining Raman spectroscopy and single-molecule microscopy. Flippable Mirrors (FM) allow to switch between pulsed Ti:Sapphire and a continuous diode-laser (\SI{640}{\nm}) excitation. Fluorescence or Raman signal, collected through a Spatial (SF) and Long-Pass Filter (LP), can be analyzed by means of an electron-multiplied Charged-Coupled-Device camera (CCD), a spectrometer (SPT), avalanche photodiodes (APD) arranged in a Hanbury-Brown-Twiss configuration (HBT), or coupled to a Time Correlated Single Photon Counting card (TCSPC) for lifetime measurements.}
\label{setup}\end{center}
\end{figure}
In the near-field range, plasmonic rulers have been proposed \cite{Jain2007, Jain2007b} and employed to measure e.g.\ nuclease activity \cite{Liu2006}, to follow dimer assembly and DNA hybridization \cite{Sonnichsen2005}. In the simplest geometry, two metal nano particles, integrated into the sample as a probe, interact with each other, yielding a shift of the plasmonic resonance which obeys a $1/d^3$ distance dependence. Different schemes helped improving sensitivity (up to about \SIrange{10}{15}{\nm}) or maximum range (\SI{40}{\nm} \cite{Seelig2007}), by means of nanoparticle-induced lifetime modification. The promise of plasmonic rulers, however, has been partially compromised by a lack of universality, as the actual scaling laws typically depend on the nano-particle shape \cite{Tabor2008}. Our method represents a key extension of another ruler species, based on fluorescence-resonant-energy-transfer (FRET) \cite{Foerster1948, Stryer1978, Lakowicz2006}, which may be employed to measure distances well beyond \SI{10}{\nm}.

In the near field, the energy transfer rate between a donor and an acceptor dipole scales as $d^{-6}$ in free space. When the acceptor takes the form of a surface of dipoles, then integration over all possible transfer sites yields a $d^{-4}$ dependence, whereas transfer to the bulk shows a $d^{-3}$ behavior \cite{Amos1997}. Recent experiments with single emitters \cite{Tisler2013} and ensemble of emitters \cite{Gaudreau2013} have confirmed the predicted \cite{Swathi2009, Gomez-Santos2011} $d^{-4}$ distance dependence of the non-radiative transfer rate to monolayer graphene. The magnitude of such coupling, enhanced with respect to other lossy materials, is described by universal parameters (such as the fine structure constant) so that the relative distance of an object --- in particular that of a fluorescent molecule --- could be accurately determined (within few \si{\nm}), by measuring the emitter decay rate with respect to vacuum. An artistic view of our concept for a graphene nanoruler is depicted in the top left panel of figure \ref{setup}. At the single emitter level, a quantitative analysis for the decay rate modification was still missing to date.

\section{DBT single emitters}
Dibenzoterrylene molecules embedded in thin anthracene crystals (about \SI{40}{\nm}) have been employed as probes in this validation test for a graphene nanoruler. Such study represents likewise an exploratory experiment for the development of a graphene quantum nanoposition-sensor, as described in \cite{Muschik2013}.

The solid-state quantum system that we use combines the high oscillator strength and brightness of organic fluorescent dye molecules with the photostability of single-photon sources, such as inorganic quantum dots or color centers in diamond \cite{Toninelli2010}. As the anthracene crystalline matrix, which acts as a shield to oxygen diffusion, is only few-tens-of-nm-thick, coupling to external photonic structures is envisioned and has indeed been shown in \cite{Toninelli2010a}. Its impact on quantum-optics experiments is due to a narrow emission into the zero-phonon line around \SI{785}{\nm} which, at cryogenic temperatures, is not subject to dephasing and, as a result, is only limited to about \SI{40}{\MHz} by the excited-state lifetime \cite{Nicolet2007, Trebbia2009}. 
\begin{figure}
\begin{center}
\includegraphics[width=0.8\columnwidth]{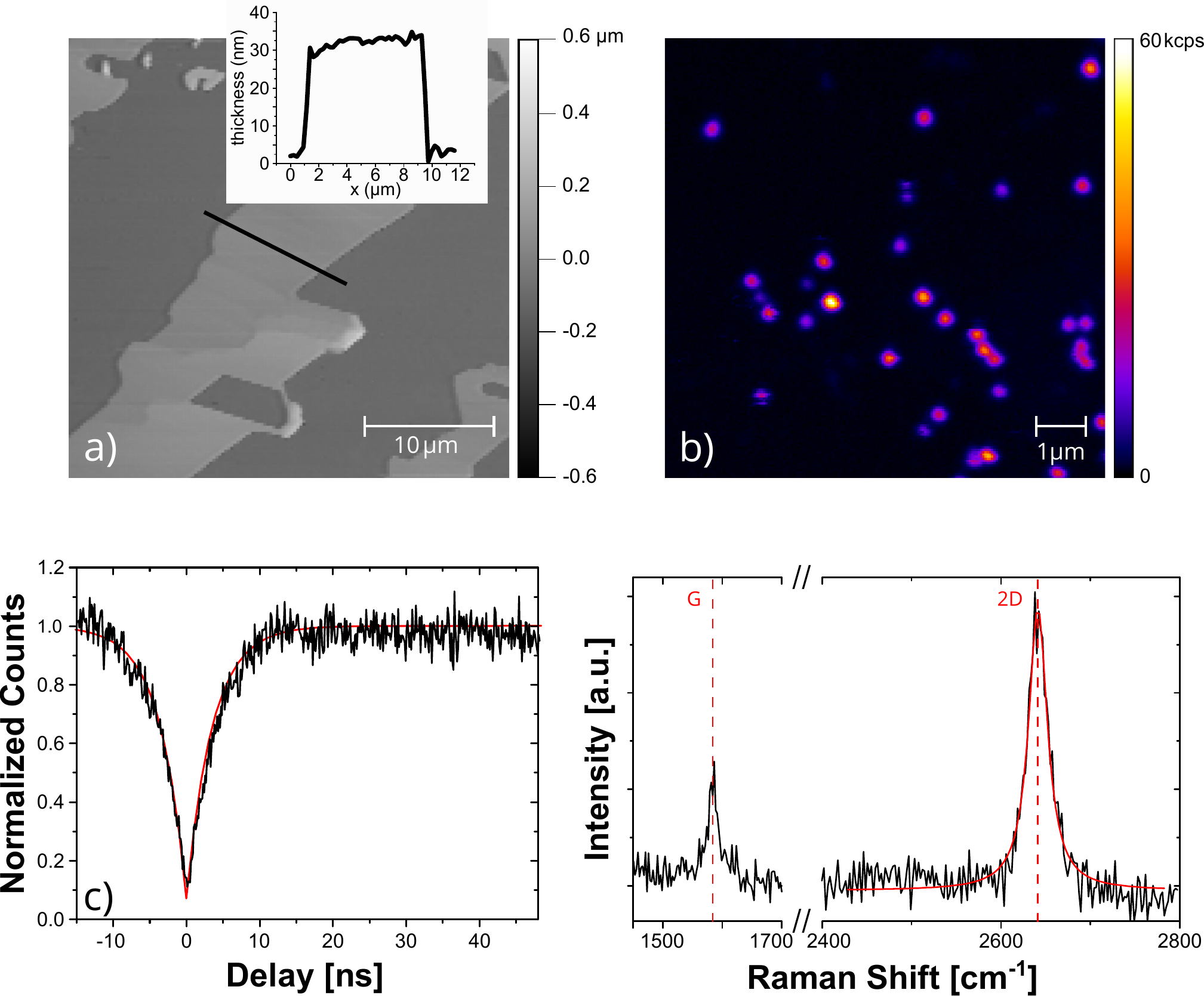}
\caption{(a) AFM image of anthracene crystals grown by spin-casting on a glass substrate. Crystal edges are clearly visible and, in the inset, a line section cut through the black line shows a thickness of \SI{30}{\nm} with low surface roughness ($\approx\SI{1}{\nm}$). (b) Confocal fluorescence scan of DBT molecules embedded in thin anthracene crystals: typical concentrations and count rates. Evidence of emission from single molecule excitation is given by measurements of the intensity autocorrelation function, such as the one displayed in panel (c). Here, photon anti-bunching from single-molecule emission is displayed in normalized units, without any background correction, and fitted with the function $ y = C (1-b\exp(-|t-t_0|/\tau)) $, yielding a coincidence reduction at $\tau=\SI{0}{\ns}$ of 93\%. (d) Raman spectrum of the CVD pristine graphene sample, obtained with a CW solid state diode laser (\SI{640}{\nm}), exhibiting the characteristic lines at $\approx\SI{1581}{\cm^{-1}}$ and $\approx\SI{2640}{\cm^{-1}}$. The single Lorentzian profile of the 2D band and the respective positions for G and 2D bands show unambiguously graphene monolayer signature.}
\label{DBTcharacterization}\end{center}
\end{figure}
For the purpose of this work, exploiting a broad-band effect, we operate at room temperature and perform a statistical analysis of  lifetimes on 150 DBT molecules, whose distances from a single graphene sheet vary from few to \SI{80}{\nm}. To fabricate the DBT-doped crystals, we use a \si{\nano\Molar} solution of DBT, prepared by mixing a solution of DBT in toluene of known concentration with a solution of anthracene in diethyl ether with a concentration of \SI{2.5}{\mg\per\ml}. The crystals are then obtained by spin-casting a \SI{20}{\micro\litre}-droplet of this solution on different substrates, including a reference glass cover slide. Following this protocol, crystals with clear-cut facets and thicknesses ranging from 20 to \SI{80}{\nm} are formed, embedding DBT at single-molecule concentration. The sample surface roughness and thickness were studied by means of Atomic Force Microscopy (AFM); a typical recorded image is shown in figure \ref{DBTcharacterization}a, displaying an average crystal thickness of \SI{30}{\nm}, and a typical surface roughness of the order of \SI{1}{\nm}. DBT is known to be hosted as an impurity in such matrices, but with a well-defined orientation, so as to minimize Gibbs free energy (figure \ref{DBTcharacterization}) \cite{Nicolet2007}. In figure \ref{DBTcharacterization}b, we show a fluorescence image of a DBT:anthracene sample deposited on silica, similar in characteristics to the one measured by AFM. The map is obtained by scanning the sample in the confocal microscope configuration, sketched in figure \ref{setup}. A diode laser at \SI{767}{\nm} (typical intensity is \SI{60}{\kW\per\square\cm}) is used in excitation and the emitted red-shifted fluorescence is collected through a 1.4-N.A. oil-immersion objective after wavelength selection. Bright spots in figure \ref{DBTcharacterization}b correspond to single molecule emission, as the second-order autocorrelation function of panel c) suggests. By reconstructing the histogram of the time intervals between start and stop events, recorded by two avalanche photodiodes (APDs) arranged in a Hanbury-Brown-Twiss geometry (see figure \ref{setup}), we approximate the intensity autocorrelation function for short time periods. We find $g^2(0)\simeq 0.2$ which is a clear indication of sub-poissonian photon statistics from a single photon source \cite{Loudon2000}. Having access to the single molecule behavior also allows for the probing of local properties such as Fermi energy spatial variation.  
In order to observe the relaxation dynamics we employ a Ti:Sapphire laser optimized to emit pulses \SI{200}{\fs} long around a wavelength of \SI{767}{\nm}, with a repetition rate of \SI{81.2}{\MHz}. The histogram of photon arrival times upon laser triggering is recorded via a PicoHarp Time-Correlated Single-Photon-Counting card. The acquired signal is well fitted by the convolution between system response function (gray curve in figure \ref{scan}) and a single exponential decay, associated to the depopulation of a single excited state. In absence of external loss channels, the quantum efficiency of our emitter is close to unity and the inverse of the population relaxation time decay is solely due to the radiative decay rate ($\simeq$\SI{25}{\MHz}). 

\section{Graphene monolayer}
The graphene monolayers under investigation have been fabricated by chemical vapor deposition (CVD) on copper.
Millimeter-size of polycrystalline graphene is then transferred on a microscope slide and annealed in H$_2$-Ar (1:5) \SI{300}{\celsius} during 3h.

The presence of a monolayer has been verified by performing Raman spectroscopy \cite{Ferrari2003} on the sample before the DBT:anthracene solution was spincasted. 
Raman spectroscopy of graphitic systems is a non-invasive, fast and complete tool to investigate doping, stress and structural properties. In particular, the two main Raman features of graphene, G and 2D bands, provide informations about the number of layers, the intrinsic graphene doping or structural defects. 
The optical setup for the analysis of the Stokes red-shifted bands is standard for confocal Raman spectroscopy, employing a 100X Objective (N.A.=0.7).  
The source is a continuous wave solid state diode laser, with central emission wavelength around \SI{640}{\nm} and TEM00 spatial mode. The excitation laser line is cleaned with a narrow (\SI{2}{\nm}) bandpass filter, whereas the Rayleigh line in detection is filtered out by means of a longpass filter. 
The spectrometer used to record Raman spectra has a spectral resolution of \SI{3}{\cm^{-1}}. As shown in figure \ref{DBTcharacterization}d, the respective positions and shape for G and 2D bands are clear signature of a graphene monolayer \cite{Saito2011, Ferrari2006, Berciaud2013}. 
The G band position is measured at \SI{1581}{\cm^{-1}} and has pure lorentzian shape. 
The profile of the Raman 2D band at \SI{2640}{\cm^{-1}} is lorentzian and shows correct agreement with the expected line shape for monolayer graphene \cite{Ferrari2006}.

\section{Lifetime measurements}

We here prove that the measurement of the decay rate is a tool to accurately pinpoint an emitter position away from a graphene interface. Specifically, we show how the lifetime distribution from a collection of 150 DBT molecules close to a graphene monolayer only depends on universal parameters, besides their position distribution. As lifetime measurements are not affected by the instrumental collection efficiency, whether geometrical or intrinsic, the system shows promising characteristics as a nanoscopic ruler. 

We consider a semi-classical model along the lines of what is described in \cite{Blanco2004} and already applied in \cite{Gaudreau2013} for the energy transfer, taking place between an ensemble of molecules and the two-dimensional material.
The decay rate of a molecule at a distance $d$ from a graphene film ($\Gamma_\mathrm{g}$) is calculated as the power radiated by a classical dipole, placed in a semi-infinite medium facing the absorbing material. $\Gamma$ is derived by integration over parallel wave vectors, considering the total field as a result of interference between the dipole emission and its Fresnel reflection. A particularly elegant form describes $\Gamma_\mathrm{g}$: in the \SI{15}{\nm}-range, using graphene DC conductivity $\sigma=e^2/4\hbar$, for a parallel oriented dipole we find:
\begin{equation}
\Gamma_\mathrm{g}/\Gamma_\mathrm{ant}\simeq1+\frac{9\alpha}{256\pi^3(\epsilon_\mathrm{ant}+\epsilon_\mathrm{sub})^2}\left(\frac{\lambda_0}{d}\right)^4
\label{transfer_eq}\end{equation}
where $\Gamma_\mathrm{ant}$ is the decay rate in the bulk medium, $\alpha$ is the fine structure constant, $\lambda_0$ is the free-space emission wavelength, $\epsilon_\mathrm{ant}$ and $\epsilon_\mathrm{sub}$ are the permittivities of the anthracene hosting medium and of the substrate supporting graphene, respectively. Besides the characteristic $d^{-4}$ dependence, inherent to the system dimensionality, the transfer mechanism appears to be well described by universal parameters, independent of the specific realization or the experimental setup. It is worth noting that the theoretical results are in agreement with a full quantum optical analysis \cite{Glauber1991}. 

\begin{figure}[h!]
\begin{center}
\includegraphics[width=0.8\textwidth]{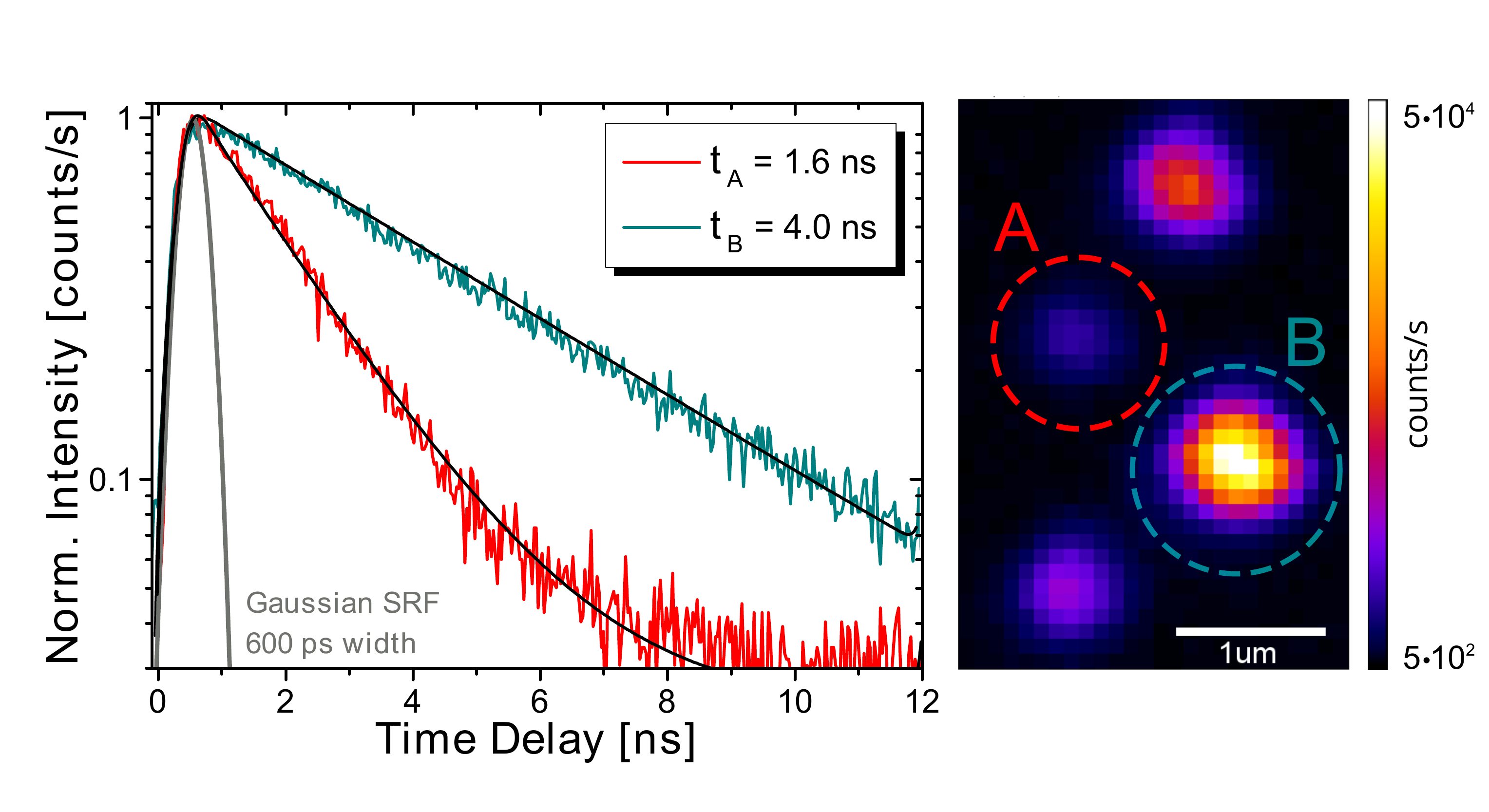}
\caption{(a) Time-resolved measurements of the fluorescence decay for single DBT molecules, placed at different distances from a graphene layer. Solid black lines are fit to the experimental data (colored on line), with the convolution between the system response function (light gray) and a single-exponential-decay function. The longer lifetime of molecule B with respect to A is associated to a brighter signal, as displayed in panel (b). Here, a close-up of a bigger confocal fluorescence scan is reported, suggesting the typical emitter concentration and signal to noise ratios.}
\label{scan}\end{center}
\end{figure}

The decay rate of molecules coupled and uncoupled to graphene have been compared in terms of excited-state lifetime measurements. The analysis was performed in a systematic way, starting from a scan similar to the one in figure \ref{DBTcharacterization}b and defining an intensity threshold for the faintest detectable molecule, according to a minimum signal to noise ratio (S/N) of 3. On each selected molecule, we then measured the relaxation dynamics and extracted a value for the excited-state lifetime. Figure \ref{scan}a shows particularly clean single-exponential decays of DBT fluorescence, in the vicinity of low-doped graphene. Such signals can be associated to a single optically active system, i.e.\ with no contribution of the host matrix, and with a simple level structure \cite{Toninelli2010}. Decay times were derived from the best fit with a convolution between the system response function (gray solid line in figure \ref{scan}) and a single-exponential decay.
The proximity to a graphene monolayer is clearly reflected in lifetime measurements: a molecule, say molecule A in the example, has been characterized by a short lifetime due to efficient energy transfer to the graphene sheet, also resulting in a fluorescence quenching. On the other hand, molecule B appears brighter and with a longer lifetime, and is supposedly further away from the graphene. Shorter lifetimes are due to the  non-radiative decay rate enhancement (expressed by equation \ref{transfer_eq}), which results in a decreased quantum efficiency, hence in fluorescence quenching.  This effect has even been exploited to image a graphene monolayer \cite{Treossi2009}, or to characterize the coupling efficiency between emitters and graphene, (see \cite{Chen2010}, quenching factor 70). 
\begin{figure}
        \centering
        \subfloat
	{
	  \includegraphics[width=0.4\columnwidth]{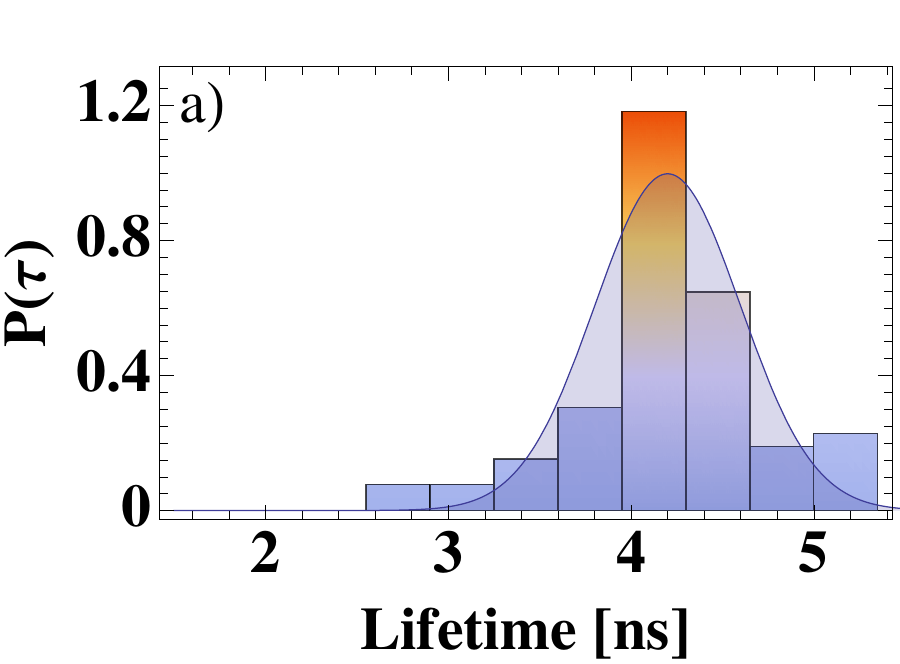}
          \label{fig:LifetimesOut}
	}
        \qquad
        \subfloat
	{
	  \includegraphics[width=0.4\columnwidth]{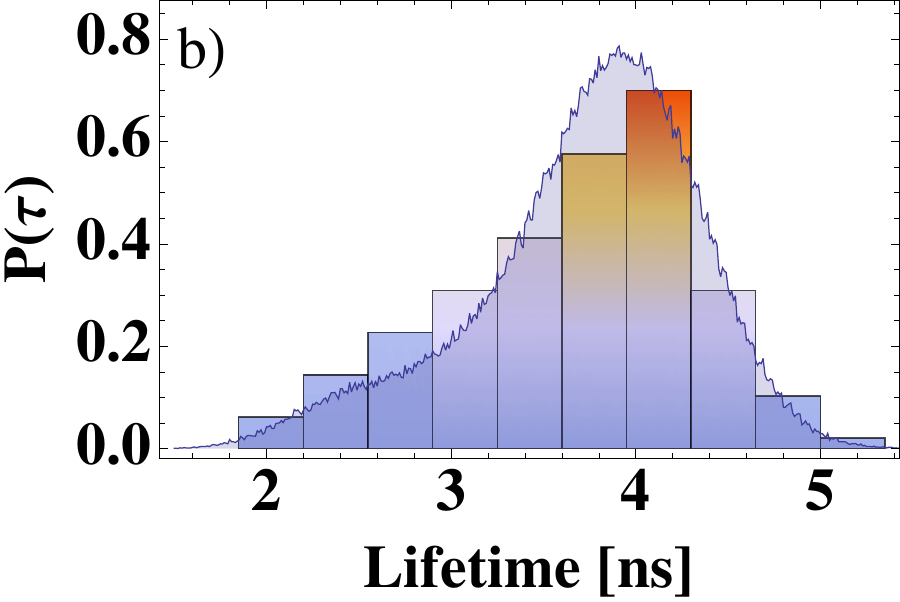}
          \label{fig:Lifetimes}
	}
	
	\subfloat
	{
	  \includegraphics[width=0.4\columnwidth]{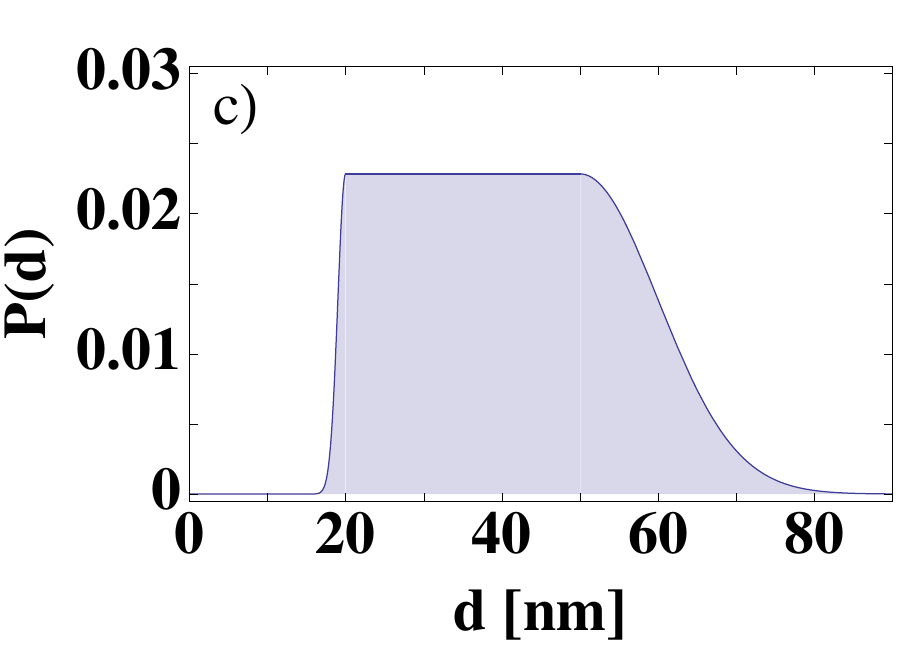}
          \label{fig:Position}
	}
	\qquad
	\subfloat
	{
	  \includegraphics[width=0.4\columnwidth]{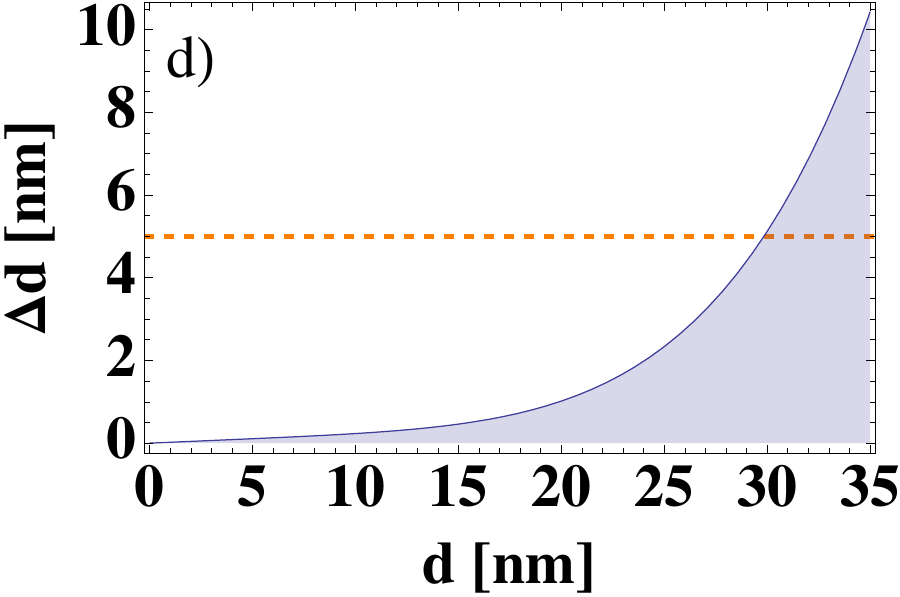}
          \label{fig:PositionUncertainty}
	}        
\caption{Probability density function (PDF) of single-molecule excited-state lifetimes on a reference sample (a) and on a graphene monolayer sheet (b). Histograms represent experimental values, while blue solid lines result from the theoretical models discussed in the text. A PDF of a normal distribution (solid line), centered around \SI{4.2}{\ns} well reproduces the intrinsic spread in lifetimes of DBT molecules in a thin anthracene film. The PDF for lifetimes on graphene requires information on the molecule position distribution, for which we assume the PDF reported in panel (c). The expected potential accuracy in operating our system to measure distances is plotted in panel (d) vs the relative distance.}
\label{Lifetimes}
\end{figure}
As a reference sample we have considered a \ce{SiO2} substrate and collected measurements for 75 molecules.
We observe therein a symmetric distribution of lifetimes around \SI{4.1}{\ns} with $\sigma\simeq \SI{0.4}{\ns}$ (see figure \ref{fig:LifetimesOut}), which cannot be fully accounted for by simple interface effects, acting differently depending on the distance to the surface. In fact, considering the in-plane orientation of DBT molecules \cite{Toninelli2010} and calculating the spread in lifetime for a \SI{40}{\nm}-thick crystal, following the discussion for a multilayer in \cite{Amos1997}, we estimate a 10\% total lifetime variation, resulting in approximately half the experimental spread. Local environment and edge effects contribute to determine the observed variation of lifetimes for DBT, as already suggested in \cite{Kreiter2002, Rogobete2004}. Such mechanisms are reasonably accounted for by a Gaussian distribution (blue solid line in figure \ref{fig:LifetimesOut}). 

When molecules are instead brought in close proximity to a mono-atomic carbon layer by spin-casting a DBT-doped anthracene thin film onto a graphene sheet, the lifetime distribution is strongly affected, becoming asymmetric with a long tail for short lifetimes. In particular, the average lifetime is shorter (\SI{3.7}{\ns}), as a new non-radiative decay channel has opened up. In the histogram shown in figure \ref{fig:Lifetimes} the bin size is given by the time resolution of the setup, amounting to \SI{400}{\ps}, after deconvolution with the instrumental response function.

In order to test the model with our actual system, we make an educated guess for the molecule position distribution, insert this assumption into eq.\ \ref{transfer_eq} and compare the obtained probability density function with the experimental data. Let us assume that DBT molecules are homogeneously distributed inside anthracene crystals, whose thicknesses are on average \SI{50\pm10}{\nm}, according to AFM measurements. A cutoff on the distribution is introduced, accounting for the before-mentioned quenching of molecules, which makes detection difficult for distances to graphene closer than \SI{20}{\nm}. In figure \ref{fig:Position}, the resulting probability density function (PDF) of the molecule position distribution is reported. As for $\Gamma_\mathrm{ant}$, we consider gaussian-distributed values, as suggested by the fit to our experimental results on the reference sample (see figure \ref{fig:LifetimesOut}). The expected PDF for the molecule lifetimes on graphene is calculated and plotted (blue solid line) for comparison to the experimental data histogram, shown in figure \ref{fig:Lifetimes}. The agreement clearly confirms the validity of the model, whose simple final formulation enables our system as a tool for position measurement at the nanoscale. In figure \ref{fig:PositionUncertainty} the uncertainty in position measurement is shown for the specific case of DBT molecules in anthracene. Here we have only included the effect of the intrinsic DBT:anthracene lifetime spread and assumed an ideal setup, i.e.\ no cutoff due to detection is considered. We observe that the distance to the graphene interface of DBT molecules can be determined with an accuracy below \SI{5}{\nm}, for distances smaller than \SI{30}{\nm}. 

In figure \ref{fig:PositionUncertainty} the uncertainty in position measurement is shown for the specific case of DBT molecules in anthracene. Here we have only included the effect of the intrinsic DBT:anthracene lifetime spread and assumed an ideal setup, i.e.\ no cutoff due to detection is considered. We observe that the distance to the graphene interface of DBT molecules can be determined with an accuracy below \SI{5}{\nm}, for distances smaller than \SI{30}{\nm}. 

Given the lifetime distribution for molecules on glass and on graphene, we can estimate the probability to measure a transfer efficiency $\eta$ higher than 40\%, which is the maximum value reported in literature for single emitters \cite{Liu2012,Tisler2013}. According to the equation $\eta=1-\frac{\tau_\mathrm{g}}{\tau_\mathrm{ref}}$, which holds when the intrinsic quantum yield amounts to 1, with $\tau_\mathrm{g}$ and $\tau_\mathrm{ref}$ being respectively the lifetime measured with and without graphene \cite{Tisler2013}, we find that in our case 12\% of the measured molecules have experienced an energy transfer efficiency higher than 40\%. Given a typical S/N for a single molecule equal to about 15 and a minimum detectable lifetime of about \SI{1}{\ns}, we do not expect (probability smaller then 0.1\%) to observe molecules with a transfer more efficient than $\simeq 70\%$. Note that such numbers are only determined by instrumental issues, such as minimum detectable lifetime and S/N. Therefore they do not represent absolute limitations for the proposed nano ruler. Finally we can estimate a maximum measured transfer efficiency from a single DBT molecule to graphene equal to $(61\pm 21)\%$. The uncertainty is estimated taking into account the fluctuations in the reference value and the precision of lifetime measurements.

\section{Conclusions}
In conclusion, we have presented a full statistical study of the coupling between single long-lived organic molecules and a graphene monolayer sheet. We have reported the highest --- to our knowledge --- ever measured transfer efficiency  from single emitters to graphene, amounting to $(61\pm 21)\%$. The molecule excited-state lifetime is strongly affected by the presence of the monoatomic carbon layer, because of its two-dimensionality, high conductivity and gapless dispersion relation. As a result, we can detect a FRET-like effect to distances well beyond the characteristic \SI{10}{\nm} of standard acceptor-donor energy transfer. The semi-classical model discussed in \cite{Gomez-Santos2011}, yielding a universal $d^{-4}$ dependence of the coupling efficiency, was successfully verified in the near-field range against a statistical distribution of the molecule lifetimes. The presented investigation on our favorable DBT:anthracene platform constitutes a first proof of principle for a graphene-based nano ruler, where ideally the distance to a surface can be measured by extracting the lifetime of a well-referenced single emitter, serving as a marker. In the near future, the use of single emitters will be essential to focus on local effects such as mapping the local Fermi energy in graphene, useful for electron-transport engineering in graphene based devices. Dibenzoterrylene molecules, emitting single photons on demand in the near infrared, are also particularly promising candidates to launch deterministic single plasmons into heavily-doped graphene.\\

This work was supported by the Seventh Framework Programme for Research of the European
Comission, under FET-Open grant MALICIA (265522) and by MIUR under FIRB Futuro in Ricerca project HYTEQ. F.\ K.\ acknowledges support by the Fundacio Cellex Barcelona, the ERC Career integration grant 294056 (GRANOP), the ERC starting grant 307806 (CarbonLight) and support by the E.\ C.\ under Graphene Flagship (contract no.\ CNECT-ICT-604391). We thank M.\ Inguscio, D.\ S.\ Wiersma, M.\ Bellini and M.\ Gurioli for fruitful discussions and continuous support. We wish to thank also B.\ Tiribilli at the ISC-CNR (Florence) for the realization of the AFM measurements and M.\ Santoro (LENS, Florence) for the Raman characterization of graphene samples.

\vspace{2 cm}

\section*{References}
\addcontentsline{toc}{section}{References}
\bibliographystyle{unsrt}  
\bibliography{graphene_ruler}

\begin{thebibliography}{10}

\bibitem{Novotny1997}
Lukas Novotny.
\newblock {Allowed and forbidden light in near-field optics. I. A single
  dipolar light source}.
\newblock {\em J. Opt. Soc. Am. A}, 14(1):91--104, 1997.

\bibitem{Moskovits1985}
Martin Moskovits.
\newblock {Surface-enhanced spectroscopy}.
\newblock {\em Rev. Mod. Phys.}, 57(3):783--826, July 1985.

\bibitem{Homola1999}
J.~Homola, Sinclair~S. Yee, and G.~Gauglitz.
\newblock {Surface plasmon resonance sensors: review}.
\newblock {\em Sensors and Actuators B: Chemical}, 54(1--2):3--15, January
  1999.

\bibitem{OBrien2009}
J.~L. O'Brien, A.~Furusawa, and J.~Vuckovic.
\newblock {Photonic quantum technologies}.
\newblock {\em Nat. Phot.}, 3(12):687--695, December 2009.

\bibitem{Drexhage1970}
K.~H. Drexhage.
\newblock {Influence of a dielectric interface on fluorescence decay time}.
\newblock {\em Journal of Luminescence}, 1--2(0):693--701, 1970.

\bibitem{Moerner1999}
W.~E. Moerner and M.~Orrit.
\newblock {Illuminating single molecules in condensed matter.}
\newblock {\em Science}, 283(5408):1670--1676, Mar 1999.

\bibitem{Michaelis1999}
J.~Michaelis, C.~Hettich, A.~Zayats, B.~Eiermann, J.~Mlynek, and V.~Sandoghdar.
\newblock {A single molecule as a probe of optical intensity distribution}.
\newblock {\em Opt. Lett.}, 24(9):581--583, May 1999.

\bibitem{Veerman1999}
J.~A. Veerman, M.~F. Garcia-Parajo, L.~Kuipers, and N.~F.~van Hulst.
\newblock {Single molecule mapping of the optical field distribution of probes
  for near-field microscopy}.
\newblock {\em Journal of Microscopy}, 194(2/3):477--482, 1999.

\bibitem{Martin-Cano2010}
D.~Mart{\'i}n-Cano, L.~Mart{\'i}n-Moreno, F.~J. Garc{\'i}a-Vidal, and
  E.~Moreno.
\newblock {Resonance Energy Transfer and Superradiance Mediated by Plasmonic
  Nanowaveguides}.
\newblock {\em Nano Lett.}, 10(8):3129--3134, July 2010.

\bibitem{tudela2011}
A.~Gonzalez-Tudela, D.~Mart{\'i}n-Cano, E.~Moreno, L.~Mart{\'i}n-Moreno,
  C.~Tejedor, and F.~J. Garc{\'i}a-Vidal.
\newblock {Entanglement of Two Qubits Mediated by One-Dimensional Plasmonic
  Waveguides}.
\newblock {\em Phys. Rev. Lett.}, 106(2):020501, January 2011.

\bibitem{tudela2014}
A.~Gonzalez-Tudela, P.~A. Huidobro, L.~Mart{\'i}n-Moreno, C.~Tejedor, and F.~J.
  Garc{\'i}a-Vidal.
\newblock {Reversible dynamics of single quantum emitters near metal-dielectric
  interfaces}.
\newblock {\em Phys. Rev. B}, 89(4):041402, January 2014.

\bibitem{Amos1997}
R.~M. Amos and W.~L. Barnes.
\newblock {Modification of the spontaneous emission rate of $\mathrm{Eu^{3+}}$
  ions close to a thin metal mirror}.
\newblock {\em Phys. Rev. B}, 55(11):7249--7254, March 1997.

\bibitem{Neto2009}
A.~H. {Castro Neto}, F.~Guinea, N.~M.~R. Peres, K.~S. Novoselov, and A.~K.
  Geim.
\newblock {The electronic properties of graphene}.
\newblock {\em Rev. Mod. Phys.}, 81(1):109--162, January 2009.

\bibitem{Koppens2011}
F.~H.~L. Koppens, D.~E. Chang, and F.~J. {Garc{\'i}a de Abajo}.
\newblock {Graphene Plasmonics: A Platform for Strong Light-Matter
  Interactions}.
\newblock {\em Nano Lett.}, 11(8):3370--3377, July 2011.

\bibitem{Chen2010}
Z.~Chen, S.~Berciaud, C.~Nuckolls, T.~F. Heinz, and L.~E. Brus.
\newblock {Energy Transfer from Individual Semiconductor Nanocrystals to
  Graphene}.
\newblock {\em ACS Nano}, 4(5):2964--2968, April 2010.

\bibitem{Zhang2011}
M.~Zhang, B.~C. Yin, X.~F. Wang, and B.~C. Ye.
\newblock {Interaction of peptides with graphene oxide and its application for
  real-time monitoring of protease activity}.
\newblock {\em Chem. Commun.}, 47(8):2399--2401, 2011.

\bibitem{Wang2011g}
Y.~Wang, Z.~Li, J.~Wang, J.~Li, and Y.~Lin.
\newblock {Graphene and graphene oxide: biofunctionalization and applications
  in biotechnology}.
\newblock {\em Trends in Biotechnology}, 29(5):205--212, May 2011.

\bibitem{Muschik2013}
C.~A. Muschik, S.~Moulieras, M.~Lewenstein, F.~H.~L. Koppens, and D.~E. Chang.
\newblock {Harnessing vacuum forces for quantum sensing of graphene motion}.
\newblock arXiv:1304.8090v1 [quant-ph].

\bibitem{Puller2013}
V.~Puller, B.~Lounis, and F.~Pistolesi.
\newblock {Single Molecule Detection of Nanomechanical Motion}.
\newblock {\em Phys. Rev. Lett.}, 110(12):125501, March 2013.

\bibitem{Arcizet2011}
O.~Arcizet, V.~Jacques, A.~Siria, P.~Poncharal, P.~Vincent, and S.~Seidelin.
\newblock {A single nitrogen-vacancy defect coupled to a nanomechanical
  oscillator}.
\newblock {\em Nat. Phys.}, 7(11):879--883, November 2011.

\bibitem{Jain2007}
P.~K. Jain and M.~A. El-Sayed.
\newblock {Universal Scaling of Plasmon Coupling in Metal Nanostructures:
  Extension from Particle Pairs to Nanoshells}.
\newblock {\em Nano Lett.}, 7(9):2854--2858, August 2007.

\bibitem{Jain2007b}
P.~K. Jain, Wenyu Huang, and M.~A. El-Sayed.
\newblock {On the Universal Scaling Behavior of the Distance Decay of Plasmon
  Coupling in Metal Nanoparticle Pairs: A Plasmon Ruler Equation}.
\newblock {\em Nano Lett.}, 7(7):2080--2088, June 2007.

\bibitem{Liu2006}
G.~L. Liu, Y.~Yin, S.~Kunchakarra, B.~Mukherjee, D.~Gerion, S.~D. Jett, D.~G.
  Bear, J.~W. Gray, A.~P. Alivisatos, L.~P. Lee, and F.~F. Chen.
\newblock {A nanoplasmonic molecular ruler for measuring nuclease activity and
  DNA footprinting}.
\newblock {\em Nat. Nanotech.}, 1(1):47--52, October 2006.

\bibitem{Sonnichsen2005}
C.~Sonnichsen, B.~M Reinhard, J.~Liphardt, and A.~P. Alivisatos.
\newblock {A molecular ruler based on plasmon coupling of single gold and
  silver nanoparticles}.
\newblock {\em Nat. Biotech.}, 23(6):741--745, June 2005.

\bibitem{Seelig2007}
J.~Seelig, K.~Leslie, A.~Renn, S.~K{\"u}hn, V.~Jacobsen, M.~van~de Corput,
  C.~Wyman, and V.~Sandoghdar.
\newblock {Nanoparticle-Induced Fluorescence Lifetime Modification as
  Nanoscopic Ruler: Demonstration at the Single Molecule Level}.
\newblock {\em Nano Lett.}, 7(3):685--689, February 2007.

\bibitem{Tabor2008}
C.~Tabor, R.~Murali, M.~Mahmoud, and M.~A. El-Sayed.
\newblock {On the Use of Plasmonic Nanoparticle Pairs As a Plasmon Ruler: The
  Dependence of the Near-Field Dipole Plasmon Coupling on Nanoparticle Size and
  Shape}.
\newblock {\em J. Phys. Chem. A}, 113(10):1946--1953, December 2008.

\bibitem{Foerster1948}
Th. Foerster.
\newblock {Zwischenmolekulare Energiewanderung und Fluoreszenz}.
\newblock {\em Ann. Phys.}, 437(1-2):55--75, 1948.

\bibitem{Stryer1978}
L~Stryer.
\newblock {Fluorescence Energy Transfer as a Spectroscopic Ruler}.
\newblock {\em Annual Review of Biochemistry}, 47(1):819--846, 1978.

\bibitem{Lakowicz2006}
Joseph~R. Lakowicz.
\newblock {\em {Principles of Fluorescence Spectroscopy}}.
\newblock Springer, 2006.

\bibitem{Tisler2013}
J.~Tisler, T.~Oeckinghaus, R.~J. St{\"o}hr, R.~Kolesov, R.~Reuter, Friedemann
  Reinhard, and J.~Wrachtrup.
\newblock {Single Defect Center Scanning Near-Field Optical Microscopy on
  Graphene}.
\newblock {\em Nano Lett.}, 13(7):3152--3156, June 2013.

\bibitem{Gaudreau2013}
L.~Gaudreau, K.~J. Tielrooij, G.~E. D.~K. Prawiroatmodjo, J.~Osmond,
  F.~J.~Garc{\'i}a de~Abajo, and F.~H.~L. Koppens.
\newblock {Universal Distance-Scaling of Nonradiative Energy Transfer to
  Graphene}.
\newblock {\em Nano Lett.}, 13(5):2030--2035, March 2013.

\bibitem{Swathi2009}
R.~S. Swathi and K.~L. Sebastian.
\newblock {Long range resonance energy transfer from a dye molecule to graphene
  has $\textup{(distance)}^{-4}$ dependence}.
\newblock {\em Journal of Chemical Physics}, 130(8):086101, February 2009.

\bibitem{Gomez-Santos2011}
G.~Gomez-Santos and T.~Stauber.
\newblock {Fluorescence quenching in graphene: A fundamental ruler and evidence
  for transverse plasmons}.
\newblock {\em Phys. Rev. B}, 84(16):165438, October 2011.

\bibitem{Toninelli2010}
C.~Toninelli, K.~Early, J.~Bremi, A.~Renn, S.~G{\"o}tzinger, and V.~Sandoghdar.
\newblock {Near-infrared single-photons from aligned molecules in ultrathin
  crystallinefilms at room temperature}.
\newblock {\em Opt. Express}, 18(7):6577--6582, March 2010.

\bibitem{Toninelli2010a}
C.~Toninelli, Y.~Delley, T.~St{\"o}ferle, A.~Renn, S.~G{\"o}tzinger, and
  V.~Sandoghdar.
\newblock {A scanning microcavity for in situ control of single-molecule
  emission}.
\newblock {\em Applied Physics Letters}, 97(2):021107, July 2010.

\bibitem{Nicolet2007}
A.~A.~L. Nicolet, P.~Bordat, C.~Hofmann, M.~A. Kol'chenko, B.~Kozankiewicz,
  R.~Brown, and M.~Orrit.
\newblock {Single dibenzoterrylene molecules in an anthracene crystal: Main
  insertion sites}.
\newblock {\em Chemphyschem}, 8:1929--1936, 2007.

\bibitem{Trebbia2009}
J.-B. Trebbia, H.~Ruf, Ph. Tamarat, and B.~Lounis.
\newblock {Efficient generation of near infra-red single photons from the
  zero-phonon line of a single molecule}.
\newblock {\em Opt. Express}, 17(26):23986--23991, 2009.

\bibitem{Loudon2000}
Rodney Loudon.
\newblock {\em {The quantum theory of light}}.
\newblock Oxford university press, 2000.

\bibitem{Ferrari2003}
G.~Ferrari, P.~Cancio, R.~Drullinger, G.~Giusfredi, N.~Poli, M.~Prevedelli,
  C.~Toninelli, and G.~M. Tino.
\newblock {Precision Frequency Measurement of Visible Intercombination Lines of
  Strontium}.
\newblock {\em Phys. Rev. Lett.}, 91(24):243002--, December 2003.

\bibitem{Saito2011}
R.~Saito, M.~Hofmann, G.~Dresselhaus, A.~Jorio, and M.~S. Dresselhaus.
\newblock {Raman spectroscopy of graphene and carbon nanotubes}.
\newblock {\em Advances in Physics}, 60(3):413--550, 2011.

\bibitem{Ferrari2006}
A.~C. Ferrari, J.~C. Meyer, V.~Scardaci, C.~Casiraghi, M.~Lazzeri, F.~Mauri,
  S.~Piscanec, D.~Jiang, K.~S. Novoselov, S.~Roth, and A.~K. Geim.
\newblock {Raman Spectrum of Graphene and Graphene Layers}.
\newblock {\em Phys. Rev. Lett.}, 97(18):187401--, October 2006.

\bibitem{Berciaud2013}
S.~Berciaud, X.~Li, H.~Htoon, L.~E. Brus, S.~K. Doorn, and T.~F. Heinz.
\newblock {Intrinsic Line Shape of the Raman 2D-Mode in Freestanding Graphene
  Monolayers}.
\newblock {\em Nano Lett.}, 13(8):3517--3523, 2013.

\bibitem{Blanco2004}
L.~A. Blanco and F.~J. {Garc{\'i}a de Abajo}.
\newblock {Spontaneous light emission in complex nanostructures}.
\newblock {\em Phys. Rev. B}, 69(20):205414, May 2004.

\bibitem{Glauber1991}
Roy~J. Glauber and M.~Lewenstein.
\newblock {Quantum optics of dielectric media}.
\newblock {\em Phys. Rev. A}, 43(1):467--491, January 1991.

\bibitem{Treossi2009}
E.~Treossi, M.~Melucci, A.~Liscio, M.~Gazzano, P.~Samor{\`i}, and V.~Palermo.
\newblock {High-Contrast Visualization of Graphene Oxide on Dye-Sensitized
  Glass, Quartz, and Silicon by Fluorescence Quenching}.
\newblock {\em J. Am. Chem. Soc.}, 131(43):15576--15577, October 2009.

\bibitem{Kreiter2002}
M.~Kreiter, M.~Prummer, B.~Hecht, and U.~P. Wild.
\newblock {Orientation dependence of fluorescence lifetimes near an interface}.
\newblock {\em The Journal of Chemical Physics}, 117(20):9430--9433, 2002.

\bibitem{Rogobete2004}
L.~Rogobete and C.~Henkel.
\newblock {Spontaneous emission in a subwavelength environment characterized by
  boundary integral equations}.
\newblock {\em Phys. Rev. A}, 70(6):063815, December 2004.

\bibitem{Liu2012}
X.~Liu, G.~Wang, X.~Song, F.~Feng, W.~Zhu, L.~Lou, J.~Wang, H.~Wang, and
  P.~Bao.
\newblock {Energy transfer from a single nitrogen-vacancy center in nanodiamond
  to a graphene monolayer}.
\newblock {\em Applied Physics Letters}, 101(23), 2012.

\end{thebibliography}

\end{document}